# Untersuchung der Wirkung von Data Storytelling auf das Datenverständnis von Dashboard-Nutzer:innen

Valeria Zitz[1], Prof. Dr. Patrick Baier[2]

**Abstract:** Mit dem zunehmenden Einsatz von Big Data und Unternehmensanalysen hat Data Storytelling als wirksames Mittel zur Vermittlung von analytischen Erkenntnissen an das Publikum an Popularität gewonnen, um die Entscheidungsfindung zu unterstützen und die Unternehmensleistung zu verbessern. Allerdings gibt es nur wenige empirische Belege für die Auswirkungen von Data Storytelling auf das Datenverständnis. Diese Studie validiert das Konzept des Data Storytelling als Konstrukt im Hinblick auf seine Wirkung auf das Datenverständnis der Nutzer. Basierend auf einer empirischen Datenanalyse zeigen die Ergebnisse dieser Studie, dass Data Storytelling-Kompetenz positiv mit der Unternehmensleistung assoziiert ist, was teilweise durch die Entscheidungsqualität vermittelt wird. Diese Ergebnisse bieten eine theoretische Grundlage für die weitere Untersuchung potenzieller Antezedenzien und Konsequenzen von Data Storytelling.

**Keywords:** Business Analytics, Business Intelligence, Dashboards, Data Storytelling, data-driven Storytelling, Data Understanding

## 1 Einleitung

In der heutigen Geschäftswelt sammeln Unternehmen riesige Datenmengen an, um Entscheidungen optimal zu treffen und damit die Unternehmensleistung zu optimieren. Der Erfolg dieses Vorhabens hängt dabei von der Kompetenz bei der Datenanalyse sowie von der Umwandlung der Ergebnisse in verwertbare Erkenntnisse ab [Da19a, Da19b].

Zu diesem Zweck setzen Unternehmen zunehmend eine Vielzahl von Business-Analytics-Lösungen ein, um aus den gesammelten Daten aussagekräftige und relevante Erkenntnisse zu gewinnen und die Entscheidungsfindung sowohl auf strategischer als auch auf operativer Ebene zu optimieren [DZ19]. Die steigende Anzahl von Unternehmen, die Big-Data-Verfahren einsetzen, hat zu einem gesteigerten Interesse an den Forschungen zum Thema Geschäftsanalytik und dessen Herausforderungen geführt [SKIW17].

Frühere Forschungsarbeiten untersuchten, wie Unternehmen von Daten und Analysen durch die positive Beziehung zwischen Business Analytics und Unternehmensleistung profitieren können [Ay16]. Auch die Auswirkungen von Business Analytics auf die Wertschöpfung [SCTD12] und auf die strategische Planung [KO19] wurden bereits untersucht. Lennerholt et al. haben mehrere Herausforderungen für Business Analytics im

[1] Hochschule Karlsruhe, Fakultät Informatik & Wirtschaftinformatik, Moltkestraße 30, 76133 Karlsruhe, valeria.zitz@gmail.com
[2] Hochschule Karlsruhe, Fakultät Informatik & Wirtschaftinformatik, Moltkestraße 30, 76133 Karlsruhe, patrick.baier@hka.de

Zusammenhang mit Berichten und Data Storytelling identifiziert: Wenn Entscheidungen auf der Grundlage dieser Berichte getroffen werden sollen, muss ein hohes Maß an Vertrauen in den Menschen, der den Bericht erstellt hat, vorhanden sein [LLS20].

Trotz der Bedeutung von Data Storytelling für Geschäftsentscheidungen [MT17] und Wissenschaftskommunikation [Zh18] ist es in der wissenschaftlichen Literatur immer noch ein Nischenthema. Zwar wird Storytelling immer wieder als wichtiges Instrument genannt und wurde sogar in den 2017 Creativity Trends des CIO Magazine aufgeführt [Ga17], dennoch sind die Auswirkungen von Data Storytelling auf das Datenverständnis der Nutzer:innen noch nicht systematisch untersucht worden. Der Nachweis, dass Data Storytelling das Datenverständnis verbessert, steht noch aus. Ohne ein tieferes Verständnis der Auswirkungen von Data Storytelling auf das Datenverständnis, das eine wichtige Motivation für den Einsatz von Data Storytelling ist, bleibt es jedoch unklar, ob diese Motivation gerechtfertigt ist.

Data Storytelling bezeichnet den Prozess, Daten in eine narrative Struktur einzubetten, um Informationen und Erkenntnisse auf eine anschauliche und überzeugende Art und Weise zu präsentieren. Dabei verbindet Data Storytelling gutes Storytelling mit Datenvisualisierung und Datenerklärung und kann als eine automatisch generierte, schriftliche Erklärung von Daten definiert werden [NS20]. Diese Form der Datenvisualisierung hat in den letzten Jahren zunehmend an Bedeutung gewonnen und wird voraussichtlich durch die jüngsten Entwicklungen im Bereich der Large Language Models (LLM) weiter an Relevanz gewinnen. Obwohl derzeitige Programme wie PowerBI diese Funktionen noch nicht basierend auf diesen mächtigen LLM implementiert haben, besteht die Erwartung, dass zukünftige Entwicklungen in diesem Bereich Data Storytelling noch effektiver machen werden.

Die vorliegende Studie trägt dazu bei, das Verständnis von Data Storytelling und seine Auswirkungen auf das Datenverständnis der Nutzer:innen zu erweitern. Die zentralen Erkenntnisse aus einem Experiment zeigen, dass der Einsatz von Data Storytelling das Datenverständnis der Proband:innen verbessert und von ihnen positiv wahrgenommen wird. Insbesondere das Verständnis von unbekannten Metriken ist höher, wenn Dashboards Data Storytelling enthalten im Vergleich zu Dashboards mit Smart Narrative oder ohne zusätzlichen Text. Um das volle Potenzial von Data Storytelling zu erforschen, werden weitere Untersuchungen mit einem integrierten Data-Mining-Test angestrebt. Zusätzlich ist es relevant, das Interesse der Nutzer:innen an dem Thema zu erfassen und das Potenzial von Data Storytelling, mehr Interesse zu wecken, zu untersuchen.

Insgesamt trägt diese Arbeit dazu bei, die Lücke in der Forschung zum Thema Data Storytelling zu schließen und einen fundierten Beitrag zum Verständnis seiner Auswirkungen auf das Datenverständnis und die Entscheidungsfindung zu liefern.

## 2 Literaturübersicht

Im Folgenden wird Data Storytelling als eine wichtige Methode zur effektiven Kommunikation von Daten definiert und dessen Rolle bei der Visualisierung von Daten erläutert. Es werden bewährte Praktiken vorgestellt, die sich auf die nahtlose Integration von Datenvisualisierung und Erzählung konzentrieren. Anschließend werden Softwarelösungen vorgestellt, die Data Storytelling als integrierte Funktion ermöglichen.

### 2.1 Data Storytelling

Data Storytelling verbindet gutes Storytelling mit Datenvisualisierung und Datenerklärung und kann als eine automatisch generierte, schriftliche Erklärung von Daten definiert werden [NS20]. Bei Data Storytelling handelt es sich um eine sich schnell entwickelnde Forschungsrichtung, die sich auf Techniken zur Verbesserung des Datenverständnisses, des Informationsausdrucks und der Kommunikation konzentriert [SSXCFWC19].

Das Ziel von Data Storytelling ist es, Daten effizienter und effektiver zu kommunizieren [MT17]. Laut Nussbauer Knaflic (2015) erfordert Data Storytelling zwei grundlegende Fähigkeiten, die bereits in der Grund- und Sekundarschulbildung vermittelt werden. Mathematik wird benötigt, um Zahlen zu erfassen, auszuwerten und zu interpretieren. Die Sprache wird dann verwendet, um Zahlen in Wörter, Wörter in Sätze und schließlich in Geschichten zu verwandeln [Nu15].

Im Folgenden werden verschiedene Ansätze zur Definition von Erfolgsfaktoren und verschiedenen Tools für Data Storytelling vorgestellt.

**Neifer et al. [NS20]:**

1. Komplexität reduzieren
2. Visualisierungsmöglichkeiten
3. Fokus auf die zentrale Geschichte
4. Anpassung der Geschichte an das Publikum
5. Die Geschichte erzählen (Struktur und Dramaturgie)
6. Die richtigen Daten auswählen
7. Gemeinsame Arbeit an der Datengeschichte
8. Das Publikum zum Handeln motivieren

**Behera und Swain (2019):**

1. Identifizierung der Betrachter:innen
2. Ziele der Betrachter:innen dokumentieren
3. Definition von KPIs um zu betrachten, überwachen und verfolgen
4. Identifizierung des einzigen Zwecks der Geschichte

**Ojo & Heravi (2018) Schemata:**

1. Zweck: Informieren, Überzeugen, Erklären
2. Medium: Browser, Mobile
3. Story-Typ: Drill-Down, Hybrid-Diashow, Interaktivität, Autor:innen-/ Leser:innen-gesteuert
4. Repräsentation: Grafik mit Anmerkungen, Video, Magazinstil, Bild, Web-App, Frage, Spiel, Grafiken
5. Interaktivität: Interaktiv, Filterung, Auswahl,Suche, Statische Graphen
6. Technologien: Javascript, Excel, HTML, Python, Illustrator, jQuery, Tableau

**Watson (2017): Bewährte Praktiken**

1. Konzentration auf die Geschichte
2. Analyse des Publikums
3. Festlegen der Rahmenbedingungen
4. Definition des Problems oder des Konflikts
5. Aufzeigen der Lösung und der Zukunft

Die Vielfalt der Ansätze und Empfehlungen von Neifer et al., Ojo & Heravi, Behera und Swain sowie Watson bereichert das Data Storytelling. Die Schwerpunkte liegen auf der Reduzierung von Komplexität und der Anpassung an das Publikum, der Auswahl von Zweck, Medium und Repräsentationsform, der Identifizierung der Betrachter:innen und der Analyse des Publikums. Watson betont zusätzlich die Festlegung von Rahmenbedingungen. Diese vielfältigen Aspekte ermöglichen eine individuelle Anpassung des Data Storytelling an die jeweiligen Anforderungen und den Kontext, um eine effektive und zielgerichtete Datenkommunikation zu erreichen und somit eine effektive und zielgerichtete Kommunikation von Daten zu gewährleisten.

### 2.2 Tools für Data Storytelling

Mehrere Studien zur Generierung natürlicher Sprache (Natural Language Generation, NLG) zeigen, dass es möglich ist, aus verschiedenen Datentypen beschreibenden Text zu erzeugen [GA07, MLSJK19]. Viele Methoden verwenden Vorlagen, um Sätze zu generieren [SW12, LLJR13], und es wurden Techniken entwickelt, die automatisch eine Vorlage anreichern [DWYL18, YSZWL20]. Neben den wissenschaftlichen Ansätzen gibt es bereits einige Softwareprodukte, die Textgeneratorfunktionen anbieten, die den Data-Storytelling-Ansätzen bis zu einem gewissen Grad ähnlich sind:

- **Microsoft PowerBI - Smart Narrative:** wandelt Visualisierungen in Texte um (auch: integriert Text mit eigenen berechneten Werten.
- **Tableau - Explain the data:** erstellt eine Übersicht über den Datensatz, einschließlich statistischer Metriken und Ausreißer.

- **Toucan Tuco - Storytelling Studio:** hilft Anwender:innen bei der Erstellung von Datengeschichten, wählt geeignete Diagramme aus und bietet ein Tool zur Erstellung von Geschichten, um den Kontext zu spezifizieren und ein Glossar zu erstellen.
- **Arria NLG:** erstellt Daten- und Visualisierungszusammenfassungen in Textform.
- **Lexio:** fasst Datensätze in Textform zusammen, bietet angeheftete Metriken und Einblicke in die Trends der Daten.
- **Yellowfin**: kombiniert Visualisierungen und Zusammenfassungen in Textform.

Im Rahmen dieser Arbeit ist es von besonderer Relevanz, sich auf Techniken zu konzentrieren, die darauf abzielen, aussagekräftige Textinhalte auf Grundlage strukturierter Daten zu generieren. Unter den verschiedenen verfügbaren Softwarelösungen hat sich PowerBI als das Tool mit der größten Nutzeranzahl etabliert. Daher wurde es für diese Studie ausgewählt, um bedeutende Datenfakten aus einem Datensatz zu extrahieren und sowohl die Interpretation der Daten als auch deren entsprechende Visualisierung zu unterstützen.

## 3 Experiment

Data Storytelling hat in jüngster Zeit an Bedeutung gewonnen, da es als effektives Mittel zur Vermittlung von analytischen Erkenntnissen und zur Verbesserung der Entscheidungsfindung in Unternehmen dient. Trotz der steigenden Popularität gibt es jedoch nur begrenzte empirische Belege für die Auswirkungen von Data Storytelling auf das Datenverständnis. Das folgende Experiment zielt darauf ab, das Konzept des Data Storytelling als Konstrukt zu validieren und seine Wirkung auf das Datenverständnis der Nutzer zu untersuchen. Durch eine umfassende empirische Datenanalyse wird gezeigt, dass eine hohe Data Storytelling-Kompetenz positiv mit der Leistung korreliert. Die Ergebnisse dieser Studie legen den Grundstein für weitere Untersuchungen potenzieller Ursachen und Auswirkungen von Data Storytelling.

### 3.1 Experimentaufbau

Im Experiment werden die drei Dashboards mit unterschiedlichen Kennzahlen nacheinander in verschiedenen Darstellungsarten bereitgestellt, welche im folgenden Unterkapitel genauer betrachtet werden. Für jedes Dashboard sollen Wissensfragen mit Hilfe des jeweiligen Dashboards beantwortet und anschließend die Nutzungserfahrung ausgewertet werden. Das Experiment beginnt mit einer Einführung und der Aushändigung eines Fragebogens, der weitere Instruktionen und die Links zu den Dashboards enthält. Im Folgenden wird der Aufbau des Fragebogens beschrieben.

Demografische Fragen werden an den Anfang der Umfrage gestellt, um später die Signifikanz zwischen den Gruppen untersuchen zu können. Die demografischen Fragen umfassen die typischen Aspekte Geschlecht, Altersgruppe, Bildung, aber auch die Selbsteinschätzung der Datenverarbeitungsfähigkeiten und die Selbsteinschätzung des Wissens über epidemiologische Metriken. Im Anschluss an die demografischen Fragen werden die drei verschiedenen Dashboard-Teile nacheinander in einzelnen Abschnitten in unterschiedlicher Darstellungsreihenfolge evaluiert.

Die Kombinationen von Komponenten und Darstellungen sind in Tab. 1 zu sehen. In der anschließenden Auswertung der Umfrage wurden die in den Wissensfragen erreichten Punkte als Punktzahl festgehalten. Die Wissensfragen zu den dargestellten Metriken stehen als Multiple-Choice-Antworten zur Verfügung und sollen Aufschluss über das Verständnis der Daten geben, das durch die jeweilige Darstellung entstanden ist.

| **Version** | **Part 1** | **Part 2** | **Part 3** |
|---|---|---|---|
| VN 1 | SN | DS | Non |
| VN 2 | SN | Non | DS |
| VN 3 | DS | Non | SN |
| VN 4 | DS | SN | Non |
| VN 5 | Non | SN | DS |
| VN 6 | Non | DS | SN |

Tab. 1: Die Kombinationen von Komponenten (Part 1 bis Part 3) und Darstellungen Smart Narrative (SN), Data Storytelling (DN) und reine Datendarstellung (Non) in den jeweiligen Versionen (VN 1 – VN 6)

### 3.2 Gezeigte Metriken und Dashboard Repräsentationen

Um die Auswirkungen von Data Storytelling auf das Verständnis von Daten zu untersuchen, wurden epidemiologische Metriken im Zusammenhang mit der Coronavirus-Pandemie (COVID-19) zur Repräsentation ausgewählt. Die Daten, Metriken und ihre Erklärungen wurden von der OurWorldInData-Website [RMR20] übernommen. Für den Versuchsaufbau wurden die ausgewählten Metriken auf drei Dashboards verteilt. Das erste Dashboard (Abb. 1) enthält durch die deutschen Medien bekannte Metriken wie COVID-19-Todesfälle, COVID-19-Fälle und die Positivrate. Das zweite Dashboard (Abb. 2) enthält Kennzahlen, die in den Medien kaum beleuchtet wurden: die COVID-19-Todesrate und die Übersterblichkeit. Das dritte Dashboard (Abb. 3) zeigt die Übersterblichkeit aufgeschlüsselt nach verschiedenen Altersgruppen.

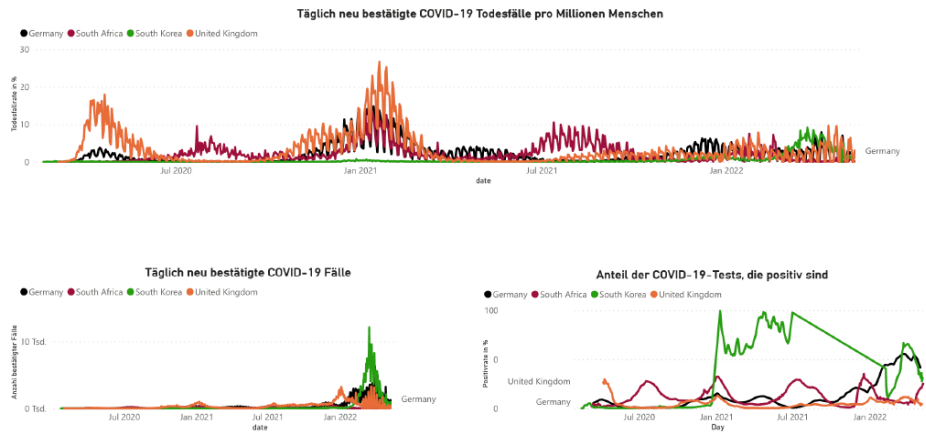

Abb. 1: Das Dashboard aus Part 1 des Experiments stellt die Kennzahlen wie COVID-19-Todesfälle, COVID-19-Fälle und die Positivrate dar.

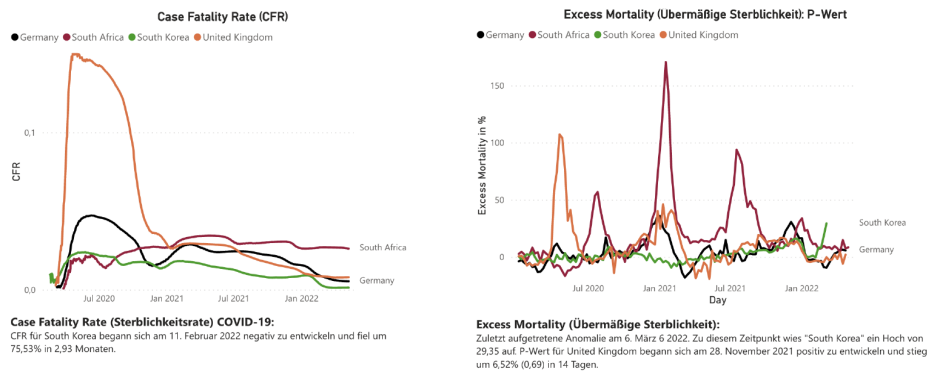

Abb. 2: Das Dashboard aus Part 2 in der Repräsentation Smart Narrative (SN) Experiments stellt die Kennzahlen Case Fatality Rate (CFR) und die Excess Mortality (Übermäßige Sterblichkeit) dar.

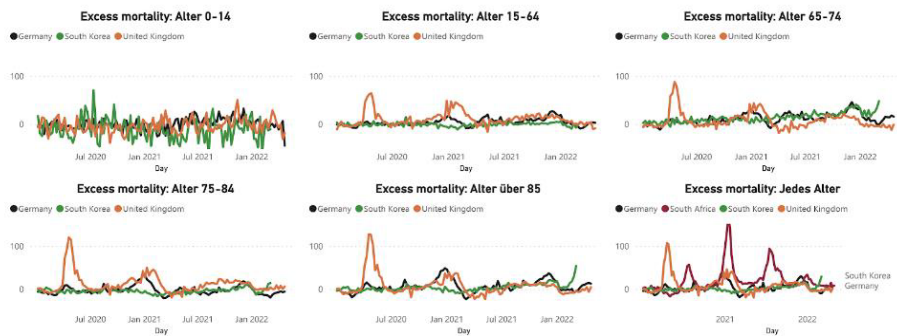

Abb. 3: Das Dashboard aus Part 3 des Experiments stellt in der Repräsentation Data Storytelling (DS) die Excess Mortality (Übermäßige Sterblichkeit) aufgeteilt in Altersgruppen dar.

Jedes der Dashboards wurde in drei verschiedenen Teilen der Umfrage mit unterschiedlichen Darstellungen präsentiert: Dashboards ohne zusätzlichen Text, wie in Abb. 1 dargestellt, enthalten nur die beschrifteten Diagramme ohne erklärende Texte. Dashboards mit Smart Narrative, wie in Abb. 2 dargestellt, stellen die beschrifteten Diagramme dar, die mit automatisch generierten Texten ergänzt wurden, die durch die Funktion Smart Narrative in Power BI (Microsoft) erstellt wurden. Abb. 3 zeigt die beschrifteten Diagramme, die mit von Menschen geschriebenen Texten ergänzt wurden. Diese Texte wurden von OurWorldInData [RMR20] übernommen und in das Dashboard als Data Storytelling entsprechend den Erfolgsfaktoren integriert.

## 4 Ergebnisse

Der folgende Abschnitt stellt die signifikanten Effekte dar, die in den Daten gefunden wurden. In den Fällen, in denen signifikante Unterschiede zwischen den verschiedenen Gruppen gefunden wurden, wird das Ergebnis angegeben. In den Fällen, in denen keine signifikanten Unterschiede gefunden wurden, werden diese aus Platzgründen nicht erwähnt. Anschließend werden die Einschränkungen der Studie diskutiert.

## 4.1 Signifikante Effekte

**Vergleich der verschiedenen Repräsentationstypen hinsichtlich des Datenverständnisses**

Die erzielte Punktzahl, die das Datenverständnis widerspiegelt, ist signifikant höher, wenn die Proband:innen die Dashboards mit Data Storytelling nutzen, im Vergleich zu den anderen Darstellungsformen. In Abb. 4 (a) wird sowohl die Gesamtpunktzahl als auch die Punktzahl in Teil 2 nach der Art der Repräsenttation dargestellt. Es ist anzumerken, dass die Teilnehmer:innen maximal 7 Punkte in den Wissensfragen erreichen konnten.

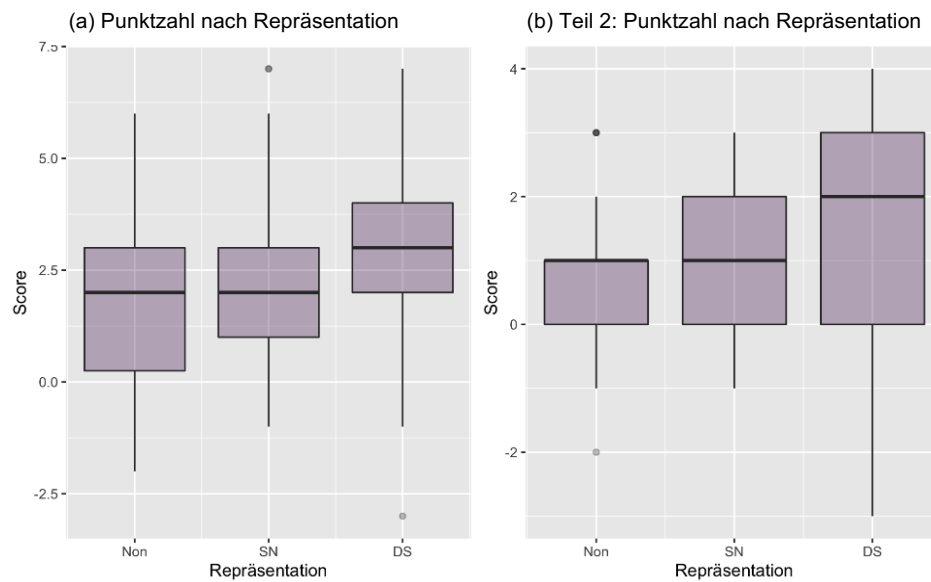

Abb. 4: Erreichte Punkzahlen der Proband:innen nach Repräsentation (ohne zusätzlichen Text (Non), Smart Narrative (SN) und Data Storytelling (DS)) aufgeschlüsselt: linke Abbildung zeigt alle Punktzahlen (a) und rechte Abbildung zeigt Punktzahlen aus Teil 2 (b)

Während zwischen den Repräsentationen in Teil 1 und Teil 3 keine signifikanten Unterschiede hinsichtlich der Punktzahlen bestand, konnten in Teil 2 signifikante Effekte hinsichtlich der Punktzahl festgestellt werden, wie in Abb. 4 (b) zu sehen ist. Demnach schneiden die Proband:innen bei der Nutzung von Data Storytelling im Durchschnitt besser ab als mit den anderen Darstellungen.

**Vergleich der verschiedenen Bildungsniveaus hinsichtlich des Datenverständnisses**

Die Teilnehmer:innen wurden in drei Gruppen eingeteilt, basierend auf ihrem Bildungsniveau und ihrem akademischen Ausbildungsbereich, wobei auch die Nähe des Ausbildungsbereichs zum Thema Daten berücksichtigt wurde. Diese Gruppierung ermöglichte eine differenzierte Analyse der Ergebnisse. Während es keinen signifikanten Unterschied zwischen Proband:innen mit unterschiedlichen Bildungsniveau in Bezug auf die erreichte Punktzahl gab, wurden bei Betrachtung der Daten aller Teile signifikante Effekte für Teil 1 und für Teil 3 gefunden, wie in Abb. 5 zu sehen ist. Demnach schneiden die Proband:innen mit einem hohen Bildungsniveau im Bereich Data Mining in Teil 1 und 3 des Experiments im Durchschnitt besser ab als mit einem niedrigen Bildungsniveau im Data Mining.

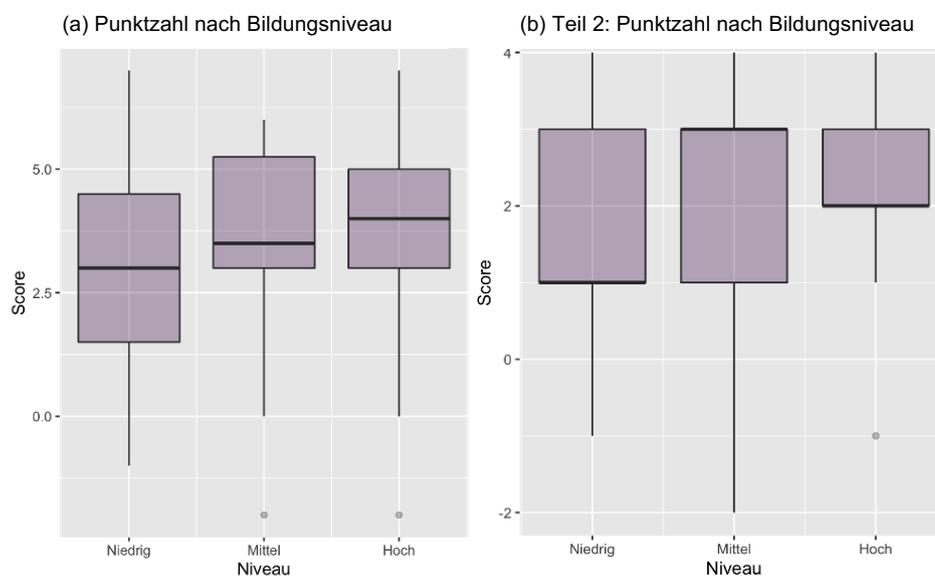

Abb. 5: In Teil 1 (a) bzw. in Teil 3(b) erreichte Punktzahl der Proband:innen aufgeschlüsselt nach den verschiedenen Bildungsniveaus der Proband:innen (Niedrig, Mittel, Hoch)

**Vergleich der verschiedenen Bildungsniveaus in Bezug auf ihre Präferenz für Dashboard-Darstellungen**

Die Ergebnisse in Abb. 6 verdeutlichen signifikante Unterschiede zwischen den Proband:innen mit unterschiedlichen Präferenzen hinsichtlich ihrer Datenverarbeitungsfähigkeiten. Es zeigt sich, dass die Mehrheit der Proband:innen mit mittlerem und hohem Bildungsniveau eine deutliche Präferenz für Data Storytelling aufweist im Vergleich zu den meisten Proband:innen mit einem niedrigeren Bildungsniveau. Dies könnte darauf hinweisen, dass Teilnehmer:innen mit einem höheren Bildungsniveau möglicherweise eine größere Affinität und Vertrautheit mit Datenvisualisierungstechniken und -methoden haben. Ihre umfassendere Ausbildung und Erfahrung im Umgang mit Daten könnten sie dazu befähigen, die Vorteile und den Wert von Data Storytelling besser zu erkennen und zu schätzen. Es ist auch möglich, dass Proband:innen mit einem niedrigeren Bildungsniveau weniger vertraut mit den Möglichkeiten von Datenvisualisierung sind und daher eher zu anderen Darstellungsformen tendieren.

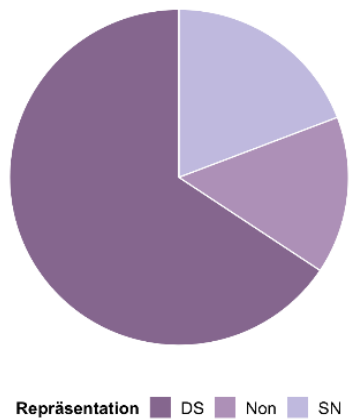

Abb. 6: Präferenzen der Proband:innen hinsichtlich der Dashboard-Repräsentationen ohne zusätzlichen Text (Non), Smart Narrative (SN) und Data Storytelling (DS)

## 4.2   Limitations

Da im Rahmen des Experiments keine eindeutige Prüfung der Kenntnisse im Bereich Data Mining stattfand, sondern das Niveau auf der Grundlage von Bildung und Selbsteinschätzung eingestuft wurde, kann dies einen Einfluss auf das Egrebnis haben. Da die untersuchten Gruppen ein überdurchschnittliches Interesse an Technik haben, ist nicht klar, ob die gezogenen Schlussfolgerungen für die gesamte Gesellschaft gelten. Da das Interesse der Proband:innen am Thema epidemiologische Metriken variieren kann, ist nicht klar, ob die Ergebnisse auf Themen übertragbar sind, die auf größeres Interesse und eine positivere Einstellung stoßen, was sich positiv auf das Verständnis für bisher unbekannte Metriken auswirken könnte.

## 5   Fazit

Die vorliegende Studie liefert starke Hinweise darauf, dass der Einsatz von Data Storytelling tatsächlich das Datenverständnis der Nutzer:innen unterstützt und von den meisten Teilnehmer:innen des Experiments positiv wahrgenommen wird. Es wurde festgestellt, dass Dashboards mit Data Storytelling ein höheres Datenverständnis für unbekannte Metriken ermöglichen im Vergleich zu Dashboards mit Smart Narrative oder ohne zusätzlichen Text.

Um das volle Potenzial von Data Storytelling weiter zu erforschen, könnten zukünftige Untersuchungen die Integration eines Data-Mining-Tests einbeziehen. Es wäre interessant, auch das Interesse der Nutzer:innen an dem Thema zu berücksichtigen und zu untersuchen, ob Data Storytelling das Potenzial hat, ein stärkeres Interesse zu wecken.

Mit der Verfügbarkeit von fortschrittlichen Sprachmodellen, wie GPT-4 und anderen, ist zu erwarten, dass Softwarelösungen in Zukunft qualitativ hochwertigere Texte generieren können, insbesondere in Bezug auf Funktionen wie Smart Narrative. Daher wäre es empfehlenswert, zukünftige Studien durchzuführen, um die Auswirkungen dieser verbesserten Textgenerierung auf das Datenverständnis und die Wahrnehmung der Nutzer:innen zu untersuchen.

Zusammenfassend lässt sich sagen, dass Data Storytelling einen positiven Einfluss auf das Datenverständnis hat und von den meisten Proband:innen als vorteilhaft empfunden wird. Zukünftige Forschung sollte sich auf die Weiterentwicklung von Textgenerierungstechnologien konzentrieren und deren Auswirkungen auf Data Storytelling untersuchen.

## 6  Literaturverzeichnis